\documentclass{article}
\usepackage{epsfig}
\usepackage{amsmath}
\usepackage{amsfonts}
\usepackage{amssymb}
\usepackage{euscript}
\usepackage{graphicx}

\begin{document}
\sloppy

{{\Huge
Does one need to consider superconductivity as Bose-Einstein condensation?
\\}}

\begin{center}
{\Large
\itshape{ B.V.Vasiliev}}
\end{center}

\section{Introduction}

The phenomenon of the Bose-Einstein condensation has some peculiarities.

    This condensate is an ordered state of bosons. Therefore, the evaporation of condensate into the normal state should be classified as the order-disorder transition with its characteristic features.

    In addition, the condensate of electrically charged particles can be destroyed both by heating and by the application of a sufficiently strong magnetic field. Therefore, between the critical temperature and the critical magnetic field of the condensate in this case there is a relationship, which must manifests itself in superconductors, if the superconductivity arises from the Bose-Einstein condensation.

Let consider these questions more detail.

\section{The Bose-Einstein condensation and the critical parameters of the condensate}
\subsection{The Bose-Einstein distribution}

 In normal metals, all energy levels up to the Fermi level at $T=0$ are filled by electrons.
 In a superconductor, electrons near the Fermi level, are combined in pairs. Let us consider the two-level system in which at T=0 these pairs occupy the energy level, which is lying on $\Delta_0$ below the Fermi-energy.
 This level with energy $\mathcal{E}_F - \Delta_0$ is basic for bosons - carriers of superconductivity. It is filled by  $N_0$ particles in a singlet state with zero total momentum and zero spin.

 Additionally, there is the level with energy $\mathcal{E}_F+\Delta_0$. It is filled by $N_1$ bosons in the excited triplet state with  spin $S=1$. \footnote{The question of the pairing of electrons in the triplet state with spin 1 has been discussed in the literature repeatedly before. See, eg, \cite{Cher}.}

The population of these levels is determined by the Bose-Einstein distribution:
\begin{equation}
N_j=\frac{1}{e^{(\varepsilon_j-\mu)/kT}-1}.\label{B-E}
\end{equation}
Where $\varepsilon_j$ is energy of $j$-th level, $\mu$ is the chemical potential.

This distribution is characterized by the fact that  at $T=0$ when $\varepsilon=0$, all particles fall into the Bose-Einstein condensate at the ground state
\begin{equation}
lim_{kT\rightarrow 0}\frac{1}{e^{(-\mu)/kT}-1}=N_0.
\end{equation}
It is satisfied at the condition
\begin{equation}
\mu=-\frac{kT}{N_0},
\end{equation}
i.e. at
\begin{equation}
\mu\cong 0\label{mu0}.
\end{equation}

The destruction of superconductivity at $T\rightarrow T_c$ occurs due to the fact that bosons
$b_0$ break, turning in a pair of fermions (electrons):
\begin{equation}
b_0\rightarrow 2e.
\end{equation}
Thus, the system consisting of a superconducting condensate of bosons, is not closed. Its density depends on temperature.

\subsection{The temperature dependence of the energetic distribution of bosons}

Let us assume that at a given temperature ${T <T_{c}}$ in accordance with Eq.(\ref{B-E}), there are  $N_0$ particles on the basic level in the singlet state and $N_1$ in the excited state.

It is important to note that the level of the Fermi-energy lies exactly between the basic and excited levels of bosons, so there should be a reaction in which these bosons split into electrons:
\begin{equation}
b_0+b_1\rightleftarrows 4e.
\end{equation}
These bosons can not be included in the condensate if they are participating in this reaction, because  virtually they are fermions.
The Bose-Einstein condensate can be formed only by the particles which make up the difference between the populations of levels $N_0-N_1$.
In dimensionless form, this difference defines the order parameter:
\begin{equation}
\Psi=\frac{N_0}{N_0+N_1}-\frac{N_1}{N_0+N_1}.
\end{equation}
In the theory of superconductivity, by definition, the order parameter is determined by the value of the energy gap
\begin{equation}
\Psi=\Delta_T/\Delta_0
\end{equation}
At taking a counting of energy from the level $\varepsilon_0$, we obtain
\begin{equation}
\frac{\Delta_{T}}{\Delta_0}=\frac{N_0-N_1}{N_0+N_1}\approx\frac{e^{2\Delta_{T}/kT} -1}{e^{2\Delta_{T}/kT} +1}=th(2\Delta_{T}/kT).\label{det}
\end{equation}
Passing to dimensionless variables $\delta\equiv \frac{\Delta_{T}}{\Delta_0}$ , $t\equiv \frac{kT}{kT_c}$
и $\beta\equiv\frac{\Delta_0}{kT_c}$ we have
\begin{equation}
\delta=\frac{e^{\beta\delta/t} -1}{e^{\beta\delta/t} +1}=th(\beta\delta/t).\label{del}
\end{equation}
This equation describes the temperature dependence of the energy gap in the spectrum of bose-particles.
It coincides in form with other equations describing other physical phenomena, which are also characterized by the existence of the temperature dependence of order parameters \cite{LL},\cite{Kit}. For example, such as the concentration of the superfluid component in liquid helium or the spontaneous magnetization of ferromagnetic materials. This equation is common for all order-disorder transitions (the phase transitions of type II in the Landau classification).

\begin{figure}
\hspace{1.5cm}
\includegraphics[scale=0.5]{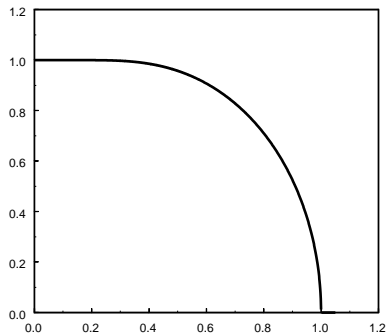}
\caption {The temperature dependence of the value of gap in the energetic spectrum of bosons calculated on Eq.(\ref{del}).}\label{D-T}
\end{figure}

The solution of this equation, obtained by the iteration method, is shown in Fig.(\ref{D-T}).
This decision is very accurately coincides with the solution of the integral equation of the BCS, which was built on the account of the phonon spectrum, and is in a quite satisfactory agreement with the measurement data.

After numerical integrating we can obtain the averaging value of the gap:
\begin{equation}
\langle\Delta\rangle=\Delta_0\int_0^1 \delta dt=0.852~\Delta_0~.\label{0.8}
\end{equation}

\subsection{The relationship between the critical parameters of boson condensate}
\subsubsection{The energy of ground state}

At the heating up to the critical temperature, the condensate particles acquire sufficient energy to break the linking between pairs.

The motion of an electron (the electric charge $e$ and mass $m_e$) with the kinetic energy  $\mathcal{E}_{kin}$ is accompanied by the magnetic field energy $\mathcal{E}_H$. At $v\ll c$ the magnetic energy \cite{Bek}:
\begin{equation}
\mathcal{E}_H\approx \alpha \mathcal{E}_{kin} \approx \alpha \frac{{m_e}v^2}{2}\label{Be}
\end{equation}
Where $\alpha=\frac{e^2}{\hbar c}$ is the fine structure constant.
Due to the fact that we are interested by the description of the behavior of electrons in metals, it is necessary to take into account that in this case, the electron mass may differ from  the free electron mass value $m_e$.

Under critical conditions, electrons have a critical velocity $v_c$ and the critical momentum
$p_c=m_e v_c$.

According to quantum mechanics, the momentum of a particle is associated with the phase of the wave function of particle $\vartheta$ with the relation:
\begin{equation}
p=\hbar \nabla \vartheta.
\end{equation}
We will assume that the condition for the condensate existence is the equation of phases of the wave functions at the boundaries of the region in which the particle is localized:
\begin{equation}
\lambda \nabla \vartheta =2\pi.
\end{equation}
Where
\begin{equation}
 \lambda=n^{-1/3},
\end{equation}
 $n$ is the pairs density into the condensate
 ($n_0=\lambda_0^{-3}$  is the density at  $T=0$).

 Under this assumption the condition of existence of the condensate would be consistent with the equality of the linear size of the volume in which the particle is localized, and its de Broglie wave:
\begin{equation}
p_c \lambda_0=2\pi\hbar.
\end{equation}

\subsubsection{The relationship between the critical parameters of a superconductor}

From the last equalities we can receive relation between the boson energy and the localization length of its particles (at $T=0$):
\begin{equation}
\Delta_0=\alpha\frac{p_c^2}{2m_2}=\frac{\alpha}{2m_2}\left(\frac{2\pi\hbar}{\lambda_0}\right)^2.\label{dl}
\end{equation}
Where  $m_2\approx 2m_e$  is the boson mass.
 Hence the particle density  of the boson condensate at $T = 0$:
\begin{equation}
n_0=\left(\frac{m_e}{\pi^2\alpha \hbar^2}\Delta_0\right)^{3/2}.\label{n0}
\end{equation}

To convert the Bose-condensate into the normal state we must give its particles (in the unit of its volume) the energy $\mathcal{E}_T$. In view of Eq.(\ref{0.8}):
\begin{equation}
\mathcal{E}_T\approx n_0 \langle\Delta_0  \rangle  \approx  0.85\left(\frac{m_e}{\pi^2\alpha\hbar^2}\right)^{3/2}\Delta_0^{5/2},\label{ET}
\end{equation}
On the other hand, we can get the normal state of an electrically charged condensate at an applying of a magnetic field with the density of energy:
\begin{equation}
\mathcal{E}_H= \frac{H_c^2}{8\pi}.\label{EH}
\end{equation}
At the equating $\mathcal{E}_T$ and $\mathcal{E}_H$, we obtain:
\begin{equation}
\Delta_0=\left(\frac{\alpha \pi^2\hbar^2}{m_e}\right)^{3/5}\left(\frac{H_c^2}{0.85 ~8\pi}\right)^{2/5}.\label{dH1}
\end{equation}

\begin{figure}
\hspace{1.5cm}
\includegraphics[scale=0.5]{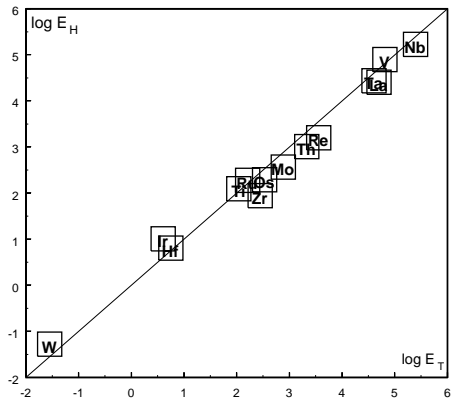}
\caption{The comparison of the critical energy densities $\mathcal{E}_T$ (Eq.(\ref{ET})) and $\mathcal{E}_H$ (Eq.(\ref{ET})) for the type II superconductors.}\label{eh-et2}
\end{figure}

The comparison of the critical energy densities $\mathcal{E}_T$ and $\mathcal{E}_H$ for type II superconductors are shown in Fig.(\ref{eh-et2}).
The obtained agreement between  energies $\mathcal{E_T}$ (Eq.(\ref{ET}))and $\mathcal{E_H}$ (Eq.(\ref{EH})) can be considered as quite satisfactory for type II superconductors ({\cite{Pool}},{\cite{Kett}}). Especially at taking into account that,
at the calculations we used the approximate value of the magnetic energy of the moving electron Eq.(\ref{Be}).
A similar comparison of data for type-I superconductors gives the results differ in several times.
 The correction  this calculation, apparently, has not  make sense.
 The purpose of these calculations was to show that the description of superconductivity as the effect of the Bose-Einstein condensation is in accordance with the available experimental data. And this goal can be considered quite reached.

\end{document}